\documentstyle[twoside,fleqn,espcrc2,epsfig]{article}

\newcommand{\AmS}{{\protect\the\textfont2
  A\kern-.1667em\lower.5ex\hbox{M}\kern-.125emS}}

\def\ltsima{$\; \buildrel < \over \sim \;$}
\def\simlt{\lower.5ex\hbox{\ltsima}}
\def\gtsima{$\; \buildrel > \over \sim \;$}
\def\simgt{\lower.5ex\hbox{\gtsima}}

\title{Seyfert Galaxies and BeppoSAX}

\author{Giorgio Matt\address{Dip. Fisica, Universit\`a degli Studi ``Roma TRE",
via della Vasca Navale 84, I-00146 Roma, Italy \\}%
        }
       
\begin{document}

\begin{abstract}
The contributions that BeppoSAX is expected to give and, after 
one and a half year of operation, has already given,
to our knowledge of both type 1 and type 2 Seyfert galaxies 
are outlined and reviewed. 
\end{abstract}

\maketitle

\section{INTRODUCTION}

In this review, rather than present and discuss in detail the first 
results obtained by BeppoSAX (see Boella et al. 1997 for an overall
description of the mission)
on Seyfert galaxies (these results are presented in detail 
elsewhere in this volume,
see in particular the contributions by Perola, Comastri and Salvati),
I will discuss current ideas on the X--ray emission of
both type 1 and 2 objects, outlining 
topics in which BeppoSAX is expected to give, and
is actually giving, important contributions. While results in the
soft and medium X--rays will be soon extended and improved by missions like
AXAF and XMM, in hard X--rays BeppoSAX results are likely to be rather lasting,
and actually unrivalled for many years. This is the reason why, in 
this contribution, I will discuss more in details issues relative to the highest
energy part of the spectrum, and more in general topics 
in which the unique broad band of BeppoSAX is particularly well suited.

\section{SETTING THE SCENARIO. THE UNIFICATION MODEL}

Let me start by describing the picture which 
is currently believed to describe Seyfert galaxies as 
a class. In this scenario, all Seyfert galaxies possess what can be
called a type--1 nucleus: a supermassive (10$^{6-9}$ M$_{\odot}$) black
hole accreting matter, probably via an accretion disc. The size of this
region is determined by the black hole gravitational radius ($R_g=GM/c^2$, 
where $M$ is the black hole mass), which ranges from 
tenths of micro-- to tenths of milli--parsecs. The nucleus
(including also the Broad Lines Region, with a size of
milli-- to tenths-- of pc) is surrounded by optically
thick matter with a roughly cylindrical symmetry (hereinafter simply called
``the torus") on at least a  pc--scale. In this unification model
(see Antonucci 1993 for a review), the type--1 nucleus is visible only
if the line--of--sight does not intercept the torus: the source is then
classified as Seyfert 1. If, on the contrary, the line--of--sight
does intercept the torus, the source is classified as Seyfert 2:
the nucleus is hidden and its presence can be argued
by indirect evidence in the optical to soft X--ray band, while it may turn
out to be directly visible in hard X--rays, when the absorbing
matter may become transparent (see sec.~\ref{sec:sey2}). 

While the unification model in its strictest version (i.e. ``the aspect
angle is the only relevant parameter") is likely to be incorrect (as it will
be discussed in  sec.~\ref{sec:umodel}),  it is certainly valid in a broad
sense. Thefeore, unless explicitly stated, in the following 
it will be assumed as the basic scenario.

\begin{figure*}[tbh]
\epsfig{file=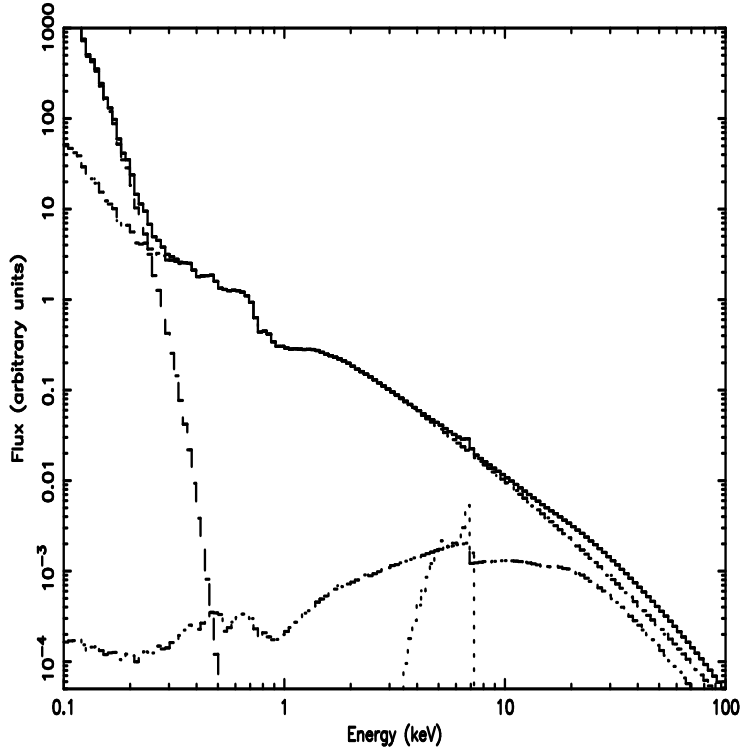, height=11.0cm, width=11.cm, angle=-90}
\caption{The X--ray spectrum of Seyfert 1s: a cut--offed
power law plus the fluorescent iron line from the relativistic disc with 
the associated reflection continuum, and a soft X--ray excess. 
The overall emission spectrum is
partly absorbed by warm matter along the line--of--sight, which left its
imprints in form of absorption edges. The figure has been produced
by using models in the {\sc xspec} code.}
\label{fig:sey1sp}
\end{figure*}

\section{SEYFERT 1}
\label{sec:sey1}

\subsection{The broad band spectrum}

Seyfert 1 galaxies have been extensively studied by all
X--ray missions, but only after GINGA it has become clear that
the spectrum is complex, resulting from different components, both in 
emission and in absorption. After  GINGA, ROSAT, ASCA and CGRO the most general 
picture is that of Fig.~\ref{fig:sey1sp}, even if not all components
shown in the figure are always and simultaneously present. 

The primary component is a power law (possibly cut--offed at high energies,
see sec~\ref{sec:emiss}), which is likely produced by Inverse Compton between
relativistic electrons and UV/soft X--ray photons coming from
the accretion disc. The details on the emission are not very well known,
despite remarkable theoretical efforts (e.g. Haardt \& Maraschi 1993;
see e.g. Svensson 1996 for a recent review, and the references therein), 
as no clear-cut signature,
able to unambiguously discriminate
between alternative models, has been detected so far. 

A significant fraction of the primary radiation is intercepted and then
reflected by circumnuclear matter, either the accretion disc or the torus,
or both. If the matter is neutral\footnote{From the hard X--rays
point of view, matter is ``neutral" as far
as Carbon and heavier elements are not fully stripped. Ionization of lighter
elements affects the spectrum only below $\approx$0.3 keV.}, 
the shape of the
reflected component is determined basically by the competition between
photoelectric absorption (whose cross section depends on the energy, after each
photoabsorption edge,  
as $\approx E^{-3}$) and Compton scattering (constant cross section, at least
up to a few tens of keV; the two cross sections are equal at about 10 keV, 
if cosmic abundances are assumed) and by
Compton downscattering. This so--called Compton reflection component
has been studied in detail in several papers (see e.g. Lightman \& White 1988;
George \& Fabian 1991; Matt, Perola \& Piro 1991);  its spectrun is 
a broad hump peaked around 30 keV. 
When added to the primary component, it hardens the total spectrum 
above a few keV and steepens it above a few tens of keV.
Besides this Compton reflection continuum, the illumination of neutral matter
by the primary radiation results also in a strong iron 6.4 keV fluorescent 
line,  emitted by iron atoms after removal of a K electron by 
an X--ray photon. 

At low energies (below $\sim$1 keV) a further component (``soft excess")
may arise. The origin of this component is rather unclear, and even ROSAT
has not been able to conclusively settle this issue. It is likely that
the soft excess is actually a mixture of different contributions which may
or may not be present simultaneously in the same source (e.g. Piro,
Matt \& Ricci 1997): the tail of thermal emission from the accretion disc
is one possibility, while
reflection from ionized matter (Ross \& Fabian 1993) is likely to 
occur in the accretion disc 
if the accretion rate is high enough (see sec.~\ref{sec:ioniz}). 

All these emission components, which are likely to originate in the vicinity
of the black hole (apart from the possible reflection component from
the torus) may pass throughout ionized matter (the ``warm absorber"), whose
main signatures are absorption 
edges of high ionization ions, mainly of oxygen atoms 
(Halpern 1984; Nandra \& Pounds 1992; Fabian et al. 1994)
In a large fraction of Seyfert 1s observed by ASCA absorption edges
have been unambiguously detected (Reynolds 1997; George et al. 1997). 
Resonant absorption lines may also be important in warm absorbers (Matt 1994; 
Krolik \& Kriss 1995; Nicastro, Matt \& Fiore 1998), and detectable by
the gratings onboard future missions (see Nicastro et al., this volume, 
for a possible ASCA detection of resonant absorption in NGC~985). 

In the following I will discuss in some detail issues related to 
the high energy part of the 
X--ray spectrum, as it is in this band that
BeppoSAX is expected to give ts best contribution, thanks to the 
unprecedented (and unrivalled also in the next future, until 
Spectrum--X--$\Gamma$,
INTEGRAL and ASTRO--E will be launched) sensitivity in hard X--rays of
the PDS instrument (Frontera et al. 1996).

\subsection{Probing the circumnuclear matter}

The two main reflectors which are supposed to 
be present around the black hole in
Seyfert galaxies are the accretion disc and the torus. Assuming that
the matter is neutral in both cases (see sec.~\ref{sec:ioniz}), 
the intensity
of the reflected flux, and the iron line equivalent width are similar
for the two reflectors (e.g. Matt et al. 1992 for the 
line equivalent widths from the accretion disc, and Ghisellini, Haardt \&
Matt 1994 and Krolik, Madau \& \.Zycki 1994
for the reflection form the torus). 
There are two ways in which the two components may be distingushed each other. 
The first is by variability studies: the reflection
component from the accretion disc should respond to 
variations in the primary
continuum on very short time scales (minutes or hours), while the torus
component should lag the primary component by years. A second and perhaps
better (at least for the impatient) possibility is to look at the iron line profile:
while the line from the torus should be narrow (i.e. unresolved by present
detectors), the line from the accretion disc is expected to be broad and
skewed owing to kinematic and 
relativistic effects (Fabian et al. 1989; Laor 1991; 
Matt et al. 1992). Such a line
has been actually detected by ASCA in the Seyfert 1 galaxy MCG--6-30-15
(Tanaka et al. 1995; see Molendi et al., this volume, for the BeppoSAX
observation of the same object): 
it was the first time, to my knowledge, that a strong--field General
Relativistic effect has been clearly observed.
Studying a large sample of objects 
observed by ASCA, Nandra et al. (1997a) have shown
that such a broad line is rather common in Seyfert 1s, even if
in no  other single source it has been so clearly detected,
due to limited
exposure times (a four days observation has been  necessary to obtain the
result on MCG--6-30-15 reported by Tanaka et al. 1995).

\subsubsection{Static or rotating black holes?}

Once the relativistic origin of the observed line broadening is established
(and no satisfactory alternatives has been proposed yet: see Fabian et al. 1995
for the discussion, and rejection, of most of them), in principle one may
hope to determine the most important disc parameters: inclination angle,
inner and outer radii. 

Most important, one could in principle also determine whether
the black hole is spinning or not. This is a key point, as one of the
most popular explanation for the radio--loud/radio--quiet dichotomy in
AGN is in term of black hole rotation, black holes in radio--quiet 
sources being static or slowly rotating, while those in radio--loud
objects being rapidly rotating (e.g. Wilson \& Colbert 1995). Seyfert
galaxies would then have static black holes, and the appropriate metric
would be the Schwarzschild one (while the Kerr metric must be used for
spinning black holes, and then for radio--loud objects). 
Kerr and Schwarzschild metrics
may be distinguished by the iron line profile. The principal difference
in the profile arises from the fact that, while in the Schwarzschild metric
the innermost stable orbit is at 6$r_g$, for a Kerr black hole it can be
as small as 1.23$r_g$ (Bardeen, Press \& Teukolsky 1972; Thorne 1974). 

\begin{figure}[tbh]
\epsfig{file=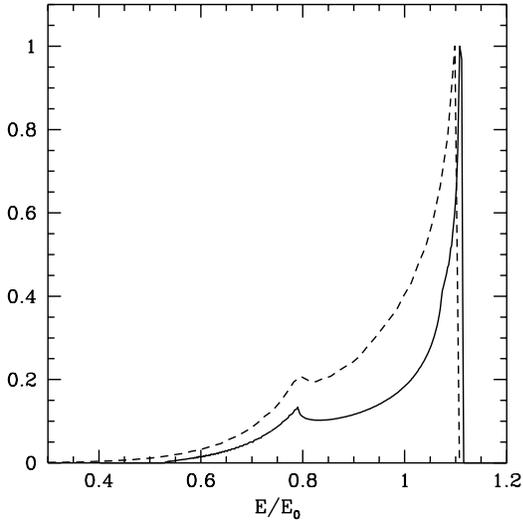, height=8.0cm, width=8.0cm }
\caption{Line profiles from  accretion discs
orbiting around static (solid curve) and extremely rotating 
(dashed curve, provided by A. Martocchia) black holes.
The inclination angle is 45$^{\circ}$, the outer radius is 20$r_g$.
A power law emissivity with index -2 is assumed. 
The inner radius is set to the last stable orbit in both cases, 
i.e. 6$r_g$ for the static black hole and 1.23$r_g$  for the rotating one.}
\label{fig:gr}
\end{figure}

In Fig.~\ref{fig:gr} the normalized line profiles 
from the accretion disc around both
static (solid curve) and maximally rotating (dashed curve)
black holes are shown. The inclination angle and the outer
radius are the same for the two profiles
(45$^{\circ}$ and 20$r_g$), while the inner radius
is set to the innermost stable orbit, i.e. 6$r_g$ and 1.23$r_g$, 
respectively. For the spinning black hole the
profile is redder than for the static black hole, the difference being
essentially due to the increased importance of the gravitational redshift
in the former case, as photons can originate from very close to the black hole.
The line emission in the two cases differs not only in shape, but
also in intensity: ``returning radiation" due to gravitational bending
(Dabrowski et al. 1997), light focusing and gravitational
blueshift of the primary radiation illuminating the disc as well as
gravitational redshift of the primary radiation escaping to the observer
(Martocchia \& Matt 1996) may increase significantly (and even dramatically,
in the Martocchia \& Matt geometry) the line intensity.

An iron line well fitted by a Kerr profile 
has been possibly observed in MCG--6-30-15
during a deep minumum phase (Iwasawa et al. 1996; see also Fabian 1997. 
This result does not
contradict the Tanaka et al. time--averaged result, as
it is possible that in the source normal state 
line emission arises from radii greater
than 6$r_g$, as Tanaka et al. have found, where differences between 
Schwarzschild and Kerr metrics are negligible). This result would 
rule out the spin of the black hole as
the key parameter for the radio--quiet/radio--loud dichotomy. However, it
is important to stress that
the main difference in the line emission between the static and spinning
black hole cases lies in the different radius of the
last stable orbit. As pointed out
by Reynolds \& Begelman (1997) the difference would be much smaller if 
efficient line emission is allowed from matter inside the last stable
orbit in the static case. This is possible if the accretion rate is high
enough to make the free--falling matter optically thick. In practice, only 
moderate accretion rates (of the order of hundredths of the Eddington value)
are required. Therefore, in such conditions Schwarzschild and Kerr black
holes are indistinguishable, at least at the first order (effects on the 
line profile due only to the different photons' 
geodesics are probably too subtle
to be testable by present and near future missions), and the 
Iwasawa et al. result would still be explanaible with a static black hole.
 A possible way to
distinguish between the two cases is by studying the ionization of the matter
(see sec.~\ref{sec:ioniz}); in fact, 
Reynolds \& Begelman (1997) also showed that the free--falling matter is likely
to be significantly ionized, independently of the accretion rate, 
while matter in a true accretion disc, may remain almost neutral provided that
the accretion rate is small enough. This point, however, has not been
fully explored yet for a Kerr black hole, and so quantitative predictions are
still lacking.

\subsubsection{Iron abundance}

Element abundances, in particular of iron, strongly affect
the reprocessed emission (see e.g. Basko 1978; George \& Fabian 1991;
Matt, Fabian \& Reynolds 1997). This occurs
in two ways: by changing  the intensity of the iron line, and by altering 
the shape of the reflected continuum. The intensity of the iron line
depends linearly on the iron abundance only in a small interval around
the solar value, deviating strongly from a linear law at very low and
very high values (see  Matt, Fabian \& Reynolds 1997 for
simple, approximated laws). The reflection continuum is affected mainly at
the iron edge, which of course deepens with increasing iron abundance 
(see Magdziarz \& Zdziarski 1995 and relative {\sc xspec} codes): between 
7.1 and, say, 20 keV  the reflection continuum is then smaller for higher
abundances. Therefore, the ratio between iron line and reflection continuum
intensity is in principle a powerful diagnostic tool for studying the iron
abundance in AGN. The combination of GINGA (Nandra \& Pounds 1994) and ASCA
(Nandra et al. 1997a) results suggests that this abundance is usually 
oversolar, a result which seems confirmed by first BeppoSAX results
(see Perola, this volume); note that BeppoSAX is the first mission
able, thanks to its unprecedented broad band,
to determine by itself this ratio with sufficient precision.

\subsubsection{Neutral or ionized matter?}
\label{sec:ioniz}

\begin{figure}[tbh]
\epsfig{file=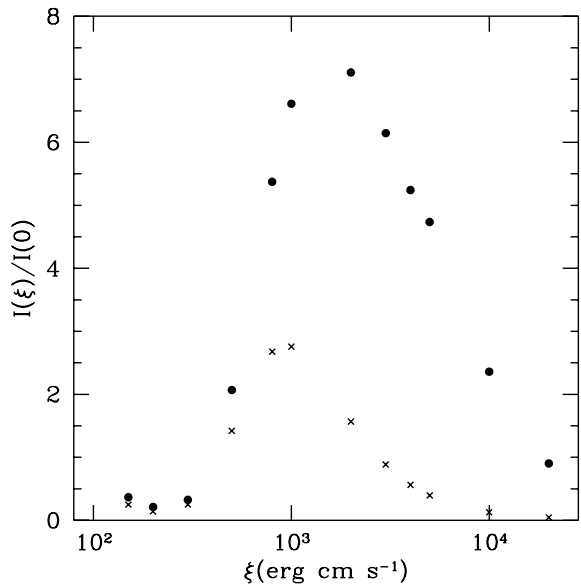, height=8.0cm, width=8.0cm }
\caption{The iron intensity (divided by the illuminating flux and 
normalized to the 
neutral matter value) as a function of the ionization parameter $\xi$. The
filled circles are the full emission, the crosses the unscattered emission
(see text). Figure adapted from Matt, Fabian \& Ross (1996).}
\label{fig:ioniz}
\end{figure}

For high enough accretion rates (i.e. 
more than a few tenths of the critical value)
the innermost accretion disc should be significantly photoionized
(see e.g. Ross \& Fabian 1993; Matt, Fabian \& Ross 1993): the ionization 
parameter $\xi$, i.e. the ratio between the flux of ionizing radiation 
and the matter density, depends in fact strongly on the accretion rate. 
The iron line intensity depends strongly on the ionization parameter: 
the fluorescent yield (i.e. the probability that a photoionization is followed
by a radiative instead than an Auger deexcitation) increases with the
ionization state of iron, and the transparency of the matter at the iron
line energy is also grater for ionized matter. However, for intermediate
ionization states, when the L shell has at least one vacancy but is still not
fully stripped, so--called Auger destruction occurs: the K$\alpha$ line
photon is resonant, and is very likely absorbed by another atom of the
next ionization state (resonant absorption cross sections are usually
order of magnitudes greater than any other relevant cross section). 
Auger deexcitation occurs 2/3 of the times, and the
photon is quickly destroyed. This mechanism does not work when iron atoms 
have the L shell completely filled, as resonant re--absorption 
is obviously impossible, 
and when the L shell is completely stripped, as there are no longer available
electrons for Auger deexcitation (and photons may eventually escape, even if 
after many resonant scatterings). The dependence of the line
intensity (divided by the illuminating flux) 
on the ionization parameter is shown 
in Fig.~\ref{fig:ioniz} (from Matt, Fabian \& Ross 1996): the decrease in the 
intensity at very high values of the ionization parameter is due to the fact
that most iron atoms are then completely stripped. It is worth noticing that
at high ionizations most photons are Compton scattered before escaping from
the matter (see Fig.~\ref{fig:ioniz}): this part of the line is very
broadened and it is probably not easy to separate it
from the underlying continuum.

As said above, iron lines from ionized matter (which are
recognizable from their highest centroid energy) are expected
in sources with  high accretion rates. ASCA observations of iron lines in
Seyfert 1s (Nandra et al. 1997) indicate that the emitting matter is
generally neutral and then that the accretion rate does not exceed about 
0.1--0.2 the critical value (according to the model of Matt, Fabian \& Ross
1993). On the other hand, high accretion rates have been suggested 
by (Laor et al. 1997) to be the basic explanation for the 
the Narrow Line Seyfert 1s phenomenon (Boller, Brandt \& Fink 1996).
In TON~S~180 BeppoSAX has actually detected such an ionized line 
(Comastri et al. 1998 and this volume), lending support to this hypothesis. 
High accretion rates and then high ionization parameters (actually so high
to have most iron atoms completely stripped) have also been invoked by 
Nandra et al. (1997b) to explain their findings, based on ASCA results, 
of an anti--correlation between iron line EW and luminosity (so--called
``X--ray Baldwin effect", originally proposed by Iwasawa \& Taniguchi 1993).

\subsection{Studying the emission mechanism}
\label{sec:emiss}

The primary emission mechanism in Seyfert 1s is not well known at present.
While Inverse Compton appears to be the basic emission process, details
are still very uncertain, and even the thermal or non--thermal nature
of the electron population is an open issue, even if the thermal hypothesis
is now more popular, after the seminal work of Haardt
\& Maraschi (1991) and the SIGMA/GRANAT 
discovery of a thermal--like cut--off in the spectrum of NGC~4151 (Jourdain
et al. 1994). Due to the limited sensitivity of hard X--ray instruments 
before BeppoSAX,  NGC~4151 has remained for many years 
the only single source in which 
such a cut--off has been unambiguously detected. BeppoSAX, thanks to the
excellent performances of the PDS, has already added one more source, namely
NGC~5548 (see Piro, this volume) to the class of the $\simlt$100 keV 
($e$--folding energy) cut--offed sources, while for other two sources,
NGC~5506 and Fairall 9, it has been able to put lower limits to 
the $e$--folding
energy definitely inconsistent with the NGC~4151 and NGC~5548 values
(see Perola, this volume). Therefore, a first result is that there is not
a universal cut--off energy (temperature?) in Seyfert galaxies.

A potentially powerful tool for studying 
the emission mechanism is broad--band spectral variability (Haardt, Maraschi
\& Ghisellini 1997), even if it usually requires long and well sampled 
observations, not easily granted by Time Allocation Committees always
dealing with heavy overbookings. 
Of course, BeppoSAX is the best suited satellite for 
this purpose, and we can be confident that in the long run it will be able
to do a good job in this respect.  

\section{SEYFERT 2}
\label{sec:sey2}

\begin{figure}[tbh]
\epsfig{file=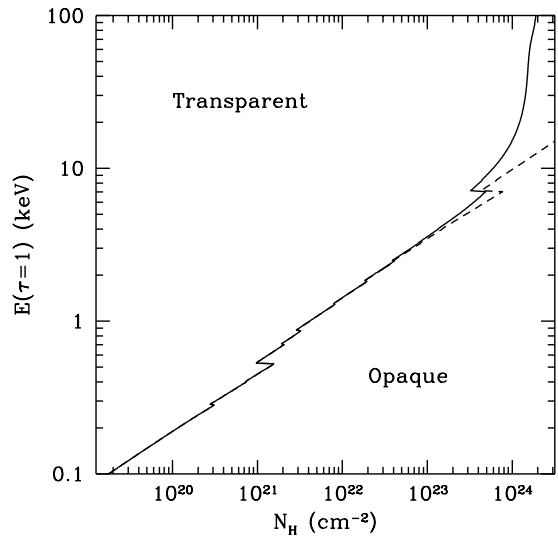, height=8.0cm, width=8.0cm }
\caption{The energy corresponding to $\tau$=1 as a function of the 
column density of the absorbing
matter. The dashed line refers to photoabsorption only, 
the solid line includes Compton scattering. }
\label{fig:nh}
\end{figure}

In the unification model, Seyfert 2 galaxies are simply Seyfert 1's 
observed throughout absorbing matter (which from now on it is  assumed to
be the molecular torus). The energy at which matter becomes
transparent, and then the appearance of the X--ray spectrum, depends
on the column density (see Fig.~\ref{fig:nh}). In particular, for $N_H$
exceeding $\sim$10$^{24}$ cm$^{-2}$, matter is optically thick at all energies
(up to the Klein--Nishina decline) owing to Compton scattering: 
the source is therefore called ``Compton--thick".
In this case, the nucleus can be observed only in scattered light, either
from optically thin, 
ionized matter\footnote{In Seyfert 1s, this matter 
may be observed in transmission rather than in reflection: i.e.
the warm reflector becomes the warm absorber discussed above.}
 or from the inner surface of the torus, or both
(Ghisellini, Haardt \& Matt 1994; Matt, Brandt \& Fabian 1996;
Matt 1996 and references
therein). The reflection from the inner surface of the torus has
the shape discussed in sec.~\ref{sec:sey1}.
 One of the signatures of the reflection from the torus 
is the 6.4 keV fluorescent iron line, with a characteristic equivalent width
(with respect to the reflected continuum) of about 1 keV: the Circinus Galaxy  
is perhaps the most spectacular example of such a line (Matt et al. 1996). 
Reflection from highly ionized matter,
on the other hand, maintains approximately the shape of the illuminating 
continuum, with superimposed fluorescent/recombination and resonant
scattering lines with equivalent widths (with respect to the
continuum reflected by the same matter) ranging from hundreds of eV
(if the matter is optically thick to resonant absorption) up to several keV
(if optically thin to resonant absorption; Matt, Brandt \& Fabian 1996).

\subsection{Mirroring the nucleus. The case of NGC~1068}

\begin{figure*}[tbh]
\epsfig{file=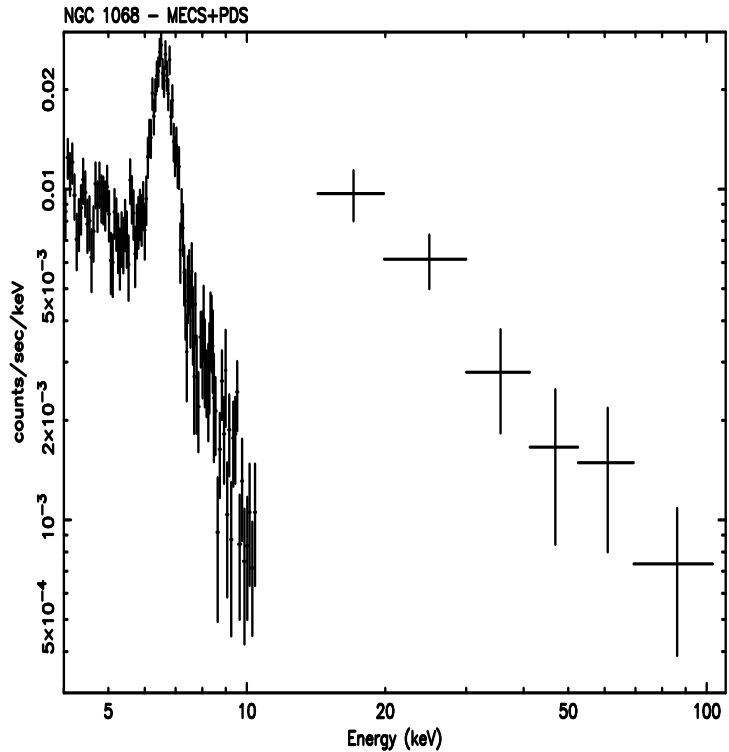, height=11.0cm, width=11.0cm, angle=-90 }
\caption{The BeppoSAX MECS+PDS spectrum of NGC~1068.}
\label{fig:1068}
\end{figure*}

NGC~1068 is the best known and most studied among Compton--thick Seyfert 2s,
and the best example of how circumnuclear matter can be probed in
these sources. BBXRT (Marshall et al. 1993) and ASCA (Ueno et al. 1994; 
Iwasawa, Fabian \& Matt 1997) have already shown, based on the presence
of both neutral and ionized iron lines, that both reflectors are at
work in this source.
To disentangle the two continua, however, hard X--rays observations
are necessary. We observed NGC~1068 with BeppoSAX for about 100 ksec,
detecting it for the first time above 10 keV (Matt et al. 1997; see
Fig.~\ref{fig:1068}). The spectrum above 4 keV (below this energy a 
thermal--like component dominates) is well fitted by a two--reflectors
model (one reflector being neutral\footnote{in the sense explained in 
footnote 1: hydrogen may well be
largely ionized, as observed in the inner region of NGC~1068, see Gallimore,
Baum \& O'Dea 1997, without affecting significantly the X--ray reflection.}
the other highly ionized) plus a
broad iron line (actually a blend of different lines). The fluxes from the two 
reflectors are comparable in the MECS range, but the neutral one, being
much harder, dominates in the PDS band. It is important
to note that, assuming an X--ray luminosity of about 10$^{44}$ erg s$^{-1}$
(see discussion in Iwasawa, Fabian \& Matt 1997), the amount of
neutral reflection (if this component is attributed, as it seems natural,
to the inner
torus surface), implies a very thick torus (more than 10$^{25}$
cm$^{-2}$) viewed almost edge--on in agreement with water maser findings 
(Gallimore et al. 1997). Note that attributing
the neutral reflection component to a rather {\it ad hoc} optically thin 
material (Netzer \& Turner 1997) would rise the problem of explaining
why reflection from the torus is not observed. 

\subsection{Is the unification model correct?}
\label{sec:umodel}

The BeppoSAX results on NGC~1068, as well as on other Seyfert 2 galaxies
(see contributions to this volume by Malaguti et al., Salvati et al.,
Ueno et al.), are brilliant confirmations of theoretical
models based on the unification scenario. However, in recent years many 
observations pointed 
against type 1 and 2 Seyfert being different only for the aspect
angle: enhanced star formation,
on average,  in Seyfert 2 galaxies (Maiolino et al. 1997); different 
morphologies between galaxies hosting type 1 and 2 nuclei, those hosting
type 2 being on average more irregular (Maiolino et al. 1997, Malkan et al. 
1997); a greater dust content in Seyfert 2s
(Malkan et al. 1997); evidence
for face--on relativistic iron lines in Seyfert 2s (Turner et al. 1997 and
this volume). Clearly, aspect angle cannot be the only relevant parameter:
there must be an intrinsic difference between average properties of
Seyfert 1 and 2. Of course, the first thing one needs to verify is whether
there is a difference in nuclear properties. To do so, the best way is to
observe in X--rays (where nuclear activity dominates the emission) 
a sample of optically
selected sources. Salvati et al. (1997 and this volume) have selected 
an OIII flux--limited sample of Seyfert 2's to be observed with BeppoSAX,
in the assumption that the OIII flux is a good isotropic indicator of
Seyfert activity (this is not completely true, but OIII flux is nevertheless
probably the best one). The main result of this program is that 
all sources observed
so far have been detected (8 out of 8), with typical X--ray luminosities
exceeding those of normal galaxies. Even if not conclusive, this is
nonetheless a very strong indication that {\sl all Seyfert 2 have a type 1
nucleus}, and then that any difference between the two classes should 
be searched for in the nuclear environment. A tentative solution is as follows:
all Seyfert have a type 1 nucleus plus
circumnuclear, optically thin (to Compton scattering) 
dust lanes (see Malkan et al. 1997); only a fraction of them, however, 
have also the (Compton--thick)
molecular torus, which possibly forms preferentially
in irregular, disturbed galaxies
(which have also, probably for the same reason, an enhanced star formation
activity as well as an overall  greater dust content). If the nucleus
is freely observed, the source is a Seyfert 1. If the line--of--sight 
intercepts matter other than the torus (i.e. a dust lane, or even the
galactic disc for highly inclined galaxies) the source falls in the 
big cauldron comprising different subclasses like Compton--thin 
Seyfert 2's, intermediate Seyfert and NELG. If, finally, the line--of--sight
intercepts the torus, the source is a Compton--thick Seyfert 2 galaxy. Note
that one of the new results from BeppoSAX is that 
Compton--thick sources are a large fraction (see e.g. the contributions
to this volume by Malaguti et al., Salvati et al. and by Ueno et al.).
Therefore, in our
proposed scenario the presence of a molecular torus, even if no longer
ubiquitous, is still rather common.

\subsection{A new population of hard sources}

One of the most important results obtained so far by BeppoSAX is the discovery,
with the MECS, of many serendipitous hard sources, sometimes observed only
above a few keV. This topic 
is extensively discussed elsewhere in this volume (see contributions
by Giommi and by Fiore et al.), and therefore I will not enter 
into details here. 
A follow--up optical identification program is currently in progress, and we
do not yet know for certain the nature of these sources. 
One can, however,
guess that these sources will turn out to be highly obscured AGN. What we
already know is that they have usually 
high X--to--optical ratios, higher
than, for instance, the sources in the Salvati et al. sample mentioned above.
It is then possible that they are
the moderate--$z$ cousins of local Seyfert 2s; in fact, if one suppose that the 
discovered sources have nuclear (and then X--ray) luminosities larger than
those of local AGN, their X--to--optical ratio (the latter
band being dominated by the host galaxy light for obscured sources) can be
explained. What is important to remark is that these hard
sources, whathever they
are, are needed to explain the Cosmic hard X--ray Background (e.g.
Comastri et al. 1995; Matt 1995 and 
references therein), and actually this BeppoSAX discovery may represent
the first step towards resolving the X--ray Background at these energies.

\section*{ACKNOWLEDGEMENTS}
This work owes much to 
contributions, discussions and suggestions from many people, actually too many
to be acknowledged here individually. I would be really ungrateful,
however,  not to 
mention G.C. Perola, F. Fiore and M. Guainazzi, the latter two for their
help in data reduction and analysis, 
and all them for frequent and stimulating discussions. 
I acknowledge financial support from A.S.I.

\end{document}